# Tunable optical emissions of $Eu^{3+}$ ions enabled by pressure-driven phase transition in ZnO

C. Ianhez-Pereira[a], U. F. Kaneko[b,c], A. D. Rodrigues[a], I. S. S. de Oliveira[d], M. P. F. de Godoy[a]

[a] Departamento de Física, Universidade Federal de São Carlos, (UFSCar), 13565-905 São Carlos-SP, Brazil.
[b] Brazilian Synchrotron Light Laboratory (LNLS), Brazilian Center for Research in Energy and Materials (CNPEM), 13083-970, Campinas-SP, Brazil.
[c] São Paulo State University (UNESP), Institute of Geosciences and Exact Sciences, 13506-900, Rio Claro-SP, Brazil.
[d] Departamento de Física, Universidade Federal de Lavras, C.P. 3037, 37203-202, Lavras-MG, Brazil.

## Abstract

Controlling the optical properties of rare-earth ions in wide-bandgap semiconductors remains a major challenge in the development of next-generation photonic materials. Here, we show that external hydrostatic pressure modulates the structural characteristics of ZnO thin films and, in turn, tunes the optical emission behavior of embedded $Eu^{3+}$ ions. By combining in situ synchrotron X-ray diffraction and photoluminescence spectroscopy under high-pressure conditions with first-principles calculations, we capture a pressure-induced phase transition from the hexagonal wurtzite to the cubic rocksalt structure near 10 GPa. This transformation is accompanied by complete quenching of the $^5D_0 \rightarrow {}^7F_J$ Europium emissions near the transition threshold, followed by a partial recovery at higher pressures, likely associated with the emergence of structural disorder. Concurrently, the Stark components of the emission bands exhibit a redshift and significant broadening with increasing pressure, reflecting enhanced crystal field strength as interatomic distances decrease. Additional first-principles calculations support the observed pressure-induced shifts in the Eu-4f states and emphasize the influence of lattice symmetry on their electronic environment. These results show that hydrostatic pressure is an effective way to adjust the optical emissions of rare-earth ions by changing their symmetry and local environment, providing a basis for designing photonic devices and luminescent materials controlled by pressure.

**Introduction**

Modulating the optical properties of rare-earth ($RE^{3+}$) ions in wide-bandgap semiconductors remains a central challenge in the design of responsive photonic materials. Among these ions, trivalent europium ($Eu^{3+}$) exhibits intra-4f transitions spanning the entire visible spectrum, making it an excellent candidate for red-emitting phosphors, display technologies, and optical sensors[1–3]. These transitions are characterized by narrow linewidths, long luminescence lifetimes, and high color purity, with energies predominantly governed by intrinsic atomic properties[4]. However, pertinent studies have demonstrated that their luminescence can be significantly modulated by factors such as the local symmetry and the strength of the crystal field in the host lattice [5,6]. Zinc oxide (ZnO), a wide-bandgap semiconductor with a wurtzite-type structure under ambient conditions, has been extensively investigated as a host for rare-earth doping due to its optical transparency, chemical stability, and technological relevance [7–9].

External hydrostatic pressure has emerged as an effective tool for modulating the optical properties of rare-earth-doped materials, primarily by inducing structural phase transitions and altering crystal field environments. Despite growing interest, pressure-responsive luminescent materials remain scarce[10], especially those capable of emission tuning under moderate pressure. Some phosphors based on $\beta$-$Ca_3(PO_4)_2$-type structures, co-doped with $Eu^{2+}$ and $Mg^{2+}$, have exhibited site-dependent emission and pressure sensitivity, but significant spectral shifts were only observed above ~16 GPa[11]. Separately, extensive work has shown how lattice strain and dopant–host interactions influence 4f-level transitions in rare-earth-doped nanomaterials[12], although external pressure effects were not considered. To our knowledge, no previous study has directly correlated $Eu^{3+}$ photoluminescence with pressure-driven structural transitions in a wide-bandgap semiconductor host, particularly within the low-to-intermediate pressure range explored here.

Alterations in the three-dimensional arrangement of atoms or ions within a crystal lattice, whether induced by doping or external stimuli, continue to provide opportunities for advanced engineering materials[13]. Notably, there has been a growing interest in studying materials under non-ambient conditions (T ~ 300 K and P ~ 0.1 MPa) to investigate phase transitions and property changes arising from altered physical states[14]. In the case of ZnO, a well-documented structural phase transition from the hexagonal wurtzite to the cubic rocksalt phase occurs at pressures near 9 GPa[15,16]. This transformation significantly impacts its optical properties, including a reduction in the bandgap from approximately 3.4 eV to 2.7 eV[16].

In this study, we investigate the structural and optical evolution of $Eu^{3+}$-doped ZnO thin films under variable hydrostatic pressure, aiming to elucidate how pressure-driven lattice modifications affect 4f-level emissions. While $Eu^{3+}$ acts primarily as an optical activator, its incorporation also introduces local lattice distortions that can influence the material's response to compression. By combining in situ synchrotron X-ray diffraction and photoluminescence spectroscopy with DFT calculations, we capture a pressure-driven structural phase transition from the hexagonal wurtzite to the cubic rocksalt structure near 10 GPa. This transition is accompanied by a quenching of the $^5D_0 \rightarrow {}^7F_J$ emissions, followed by partial recovery at higher pressures. Simultaneously, we observe emission peak broadening and a pronounced redshift, indicative of an enhanced crystal field as interatomic distances decrease. DFT results reveal pressure-induced shifts in the Eu 4f-related projected density of states, particularly within the high-symmetry cubic phase. To the best of our knowledge, this is the first study to integrate in situ synchrotron X-ray diffraction and photoluminescence spectroscopy under high-pressure conditions to directly correlate $Eu^{3+}$ luminescence behavior with pressure-driven lattice symmetry changes. This combined approach reveals fundamental connections between structure and properties in rare-earth-doped semiconductors and provides a basis for developing photonic functionalities tailored by pressure.

**Sample growth**

Eu-doped ZnO thin films were deposited onto soda-lime glass substrates using the Spray-Pyrolysis (SP) technique. Zinc acetate dihydrate ($Zn(C_2H_3O_2)_2 \cdot 2H_2O$, Synth) and hydrated europium(III) acetate

($Eu(C_2H_3O_2)_3 \cdot H_2O$ , Sigma-Aldrich) were used as precursors in distilled water, mixed in the appropriate stoichiometric ratio to achieve a 2.8±0.5 at.% Eu doping level, and stirred at room temperature.

The resulting solution was sprayed onto substrates heated at 300 °C. During deposition, the substrate temperature gradually decreased to 220 °C to optimize solvent evaporation and precursor decomposition. The spray process was conducted in cycles, allowing the substrate to recover its initial temperature between each cycle to ensure uniform film formation through solvent evaporation and solute pyrolysis. The spray flow rate was maintained at 0.5 mL/min using a syringe pump, and compressed air at 0.1 MPa served as the carrier gas.

**Characterization design**

X-ray diffraction (XRD) and Photoluminescence (PL) experiments as a function of pressure were performed at the multipurpose setup of EMA beamline at Sirius synchrotron light source[17]. An X-ray beam size of 10×10 µm$^2$ was focused on the sample position through a Kirkpatrick-Baez (KB) mirror with a wavelength of 0.4859 Å. The transmission mode XRD 2D diffraction images were captured by an area detector Rayonix SX165 model and the XRD profiles were azimuthally integrated in the Dioptas software calibrated with a NIST standard $LaB_6$. A Verdi laser from Coherent with a 532 nm wavelength and focal spot of 5 µm was used to excite the PL of the sample, which was captured by a Princeton HRS300 spectrometer equipped with a diffraction grating with 150 grooves/mm and a PIXISBR-100 Peltier cooled CCD.

ZnO:Eu thin film was mechanically reduced to dimensions of tens of microns to allow placement inside a stainless-steel gasket chamber of a diamond anvil cell (DAC) together with a Chervin ruby micro sphere and neon gas as pressure gauge and pressure transmitting medium, respectively. The half 2θ angular aperture of this DAC was estimated to be smaller than 14.5°.

First-principles calculations were performed using Density Functional Theory (DFT) as implemented in the Quantum ESPRESSO package[18]. A plane-wave basis set was employed in conjunction with the projector-augmented wave (PAW) method[19]. The exchange-correlation functional was treated within the generalized gradient approximation (GGA), using the Perdew–Burke–Ernzerhof (PBE) formulation[20]. To properly account for the strongly correlated Eu 4f electrons, the DFT+U method was applied following the Dudarev approach[21], with a U value of 7.2 eV. This value was obtained from first-principles linear response calculations[22] and is consistent with previous DFT studies[23,24]. The Kohn–Sham orbitals were expanded using a plane-wave energy cutoff of 48 Ry. Atomic geometries were relaxed until the forces on each atom were below 10 meV/Å. Brillouin zone sampling was carried out using a 6 × 6 × 6 Monkhorst–Pack k-points mesh[25].

## Results and Analysis

The X-ray diffraction (XRD) series as a function of pressure is shown in Figure 1. Three color-differentiated groups of diffraction profiles are presented, corresponding to three pressure ranges. In the first group, between 0.5 and 5.0 GPa, the pink XRD profiles exhibit the h-(100), h-(002), and h-(101) reflections, associated with the hexagonal phase. A small peak around 2θ = 12.9°, possibly corresponding to the c-(200) reflection of a cubic phase, emerges in the second group of diffraction profiles, between 5.75 and 10.0 GPa. In the green set of diffractograms, reflections related to the hexagonal phase are still observed. Above this pressure range, the blue XRD profiles show a sudden increase in the intensity of the cubic peak above 10.0 GPa, with the amount of cubic phase increasing up to 14.5 GPa. The comparable intensities of the peaks in this pressure range suggest the coexistence of hexagonal and cubic phases in similar proportions.

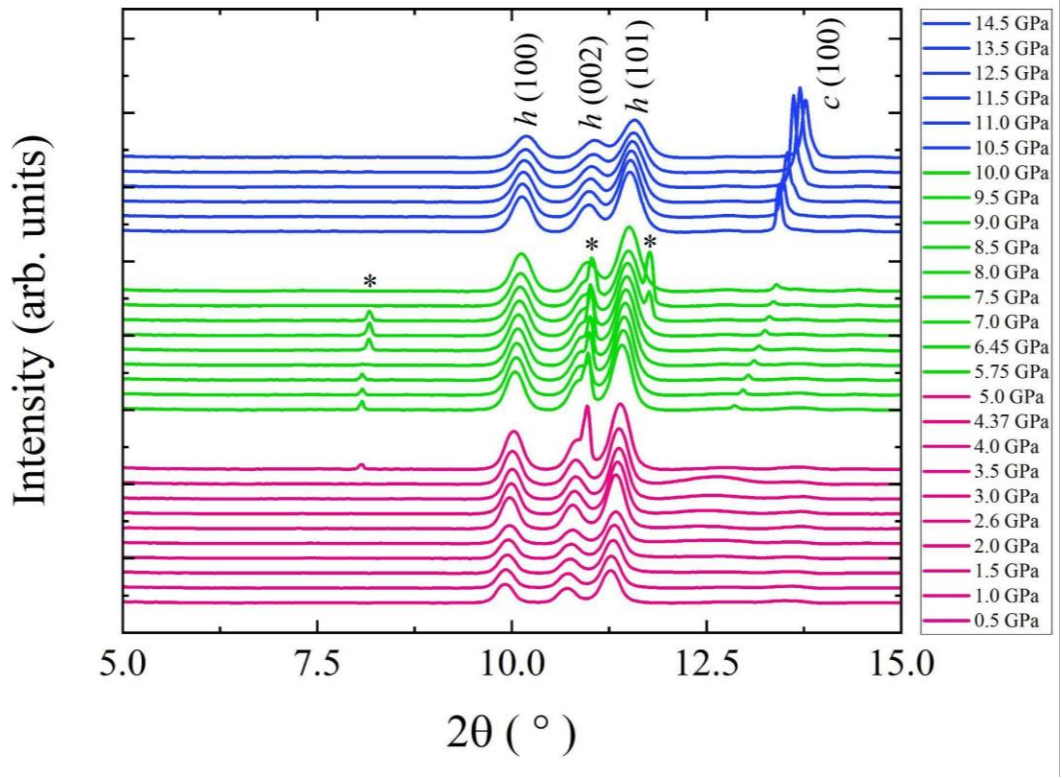

Figure 1 – Raw integrated diffractograms of ZnO:Eu as a function of pressure showing hexagonal and cubic phases. Asterisks mark extra peaks from undetermined phases.

Extra peaks appear mainly in the intermediate pressure region and are labeled with asterisks. As these peaks come from punctual structures in the 2D diffraction images, their origin is likely associated with phases with highly oriented planes, however more peaks are needed for their determination. In fact, these peaks might be masked without loss of reliability about evolution of the main phases (hexagonal and cubic) as a function of pressure.

A Rietveld refinement for the first region of this series was carried out through wurtzite structure with a space group n°186 (P6$_3$mc) in a limited angular region between 5 and 14°. Although this procedure was satisfactory for the hexagonal phase in region II, it did not work well concomitantly with the rock-salt cubic phase with space group n° 225 (Fm-3m), likely due to the vanishingly small intensity of the cubic peak (200). In this case, the angular position of the cubic peak was adjusted according to the peak angle, however, was not refined. On the contrary, the refinement employing both phases in region III showed good agreement between observed and calculated profiles from Rietveld refinements.

For pure ZnO grown by vapor deposition, a study[16] on the pressure dependence of its optical absorption edge showed that the wurtzite-to-rocksalt phase transition (w → rs) begins at approximately 9.7 GPa and is completed around 13.5 GPa, also inducing a bandgap shift from direct to indirect. Other pressure-dependent studies report that, for pure ZnO, the w → rs transition starts at P = 10.0 GPa and concludes at P = 15.0 GPa[26,27]. In the case of Eu-doped ZnO analyzed in this work, a gradual phase transition is observed. At P = 5.8 GPa, the sample appears to enter a probable polymorphic state (region II). In region III, the rs-(200) peak, associated with the cubic phase, exhibits a noticeable shift at P = 10.5 GPa, coinciding with the w → rs transition pressure reported in the literature for ZnO. The presence of small diffraction peaks at P = 5.8 GPa, in positions attributed to the rs-ZnO phase, suggests that this behavior may be influenced by dopant incorporation into the lattice or the synthesis method employed, both of which can modify the material structural characteristics. It was demonstrated that the rs-ZnO phase is metastable near ambient conditions and remains stable only at higher pressures[26], with its upper stability limit reported at ~ 56 GPa[15]. Moreover, temperature *versus* pressure phase diagrams for ZnO[28,29] indicate that the transition between hexagonal and cubic phases does not follow a well-defined pattern over a broad pressure range (approximately 2 GPa to 10 GPa), becoming more clearly established above P = 10 GPa.

The lattice parameters $a_h$ and $c_h$ of the hexagonal phase and $a_c$ of cubic phase of ZnO:Eu as a function of pressure as presented in Figure 2 (a-b) and Figure 2 (c), respectively. These graphs also display two dashed lines that separate the three pressure ranges, as shown previously in Figure 1. A continuous compression of hexagonal structure is clearly revealed by the monotonic reduction in the lattice parameters in the first two regions with small changes likely due to instabilities in the refinement procedure. Discontinuities are observed in both parameters at the onset of the region III of pressures, indicating a global structural change caused by the sudden emergence of a long-range cubic order coexisting with the preexisting hexagonal phase. Although the cubic lattice parameters in the intermediate region were obtained from the position the small c-(200) reflection, there is a good agreement in their evolution between regions II and III, as shown in Figure 2 (c), with a fast compression rate (Å/GPa) compared to the observed for the hexagonal phase in the same region.

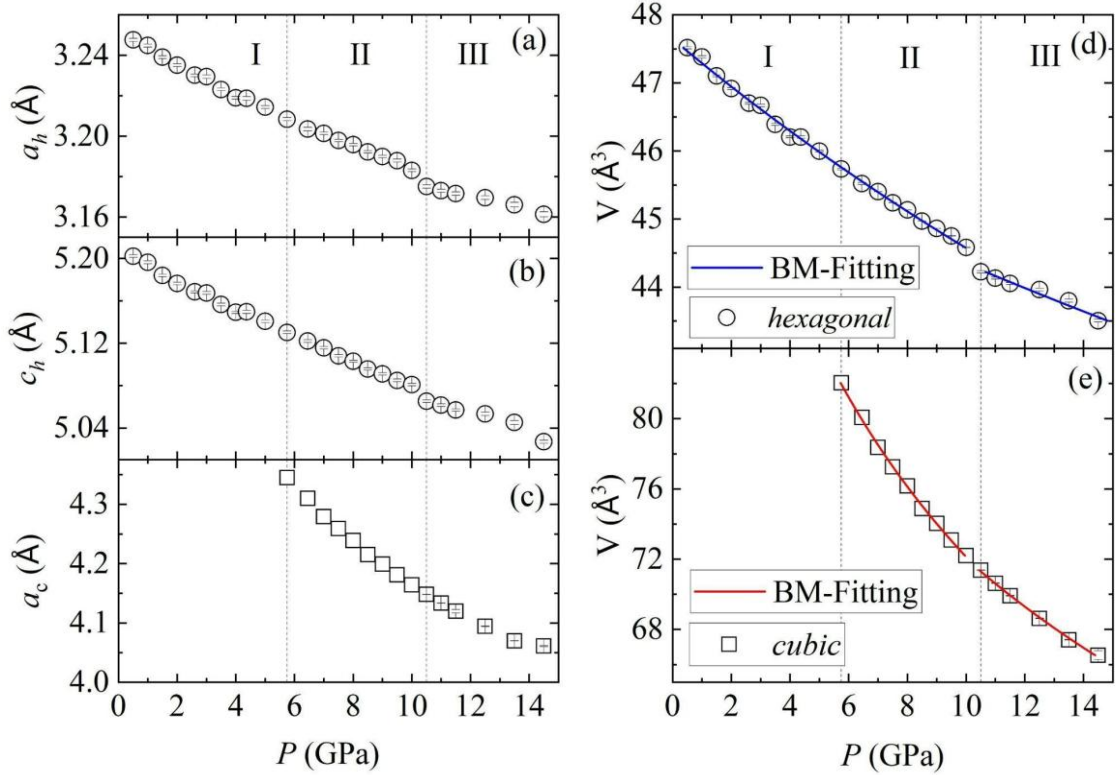

Figure 2 – Lattice parameters (a) $a_h$ and (b) $c_h$ from hexagonal phase and (c) $a_c$ from cubic phase, obtained from Rietveld Refinement of the XRD profiles of ZnO:Eu as a function of pressure and from peak position of the c-(200) in region II. Volume as a function of pressure obtained from Rietveld refinement and peak position of (d) hexagonal (black circles) and (e) cubic (red circles) phases. Red and blue solid lines are the fittings from the Birch-Murnaghan equation. Dashed lines separate regions I, II and III.

A better comprehension of the elastic response of Eu-doped ZnO is captured from the adjusting of relation between the unit cell volume and pressure through the Birch-Murnaghan (BM) equation. Figure 2 exhibits the relation between the volume and pressure obtained from the Rietveld analysis of hexagonal (d) and cubic (e) phases. In total, four fittings employing the BM equation were carried out. For the hexagonal phase (w-ZnO), the fittings come from region I + II and region III, respectively and for cubic phase (rs-ZnO) from region II and region III, respectively. The parameters obtained in each of the four fittings are organized in Table 1. Although visually the different behavior of the unit cell volume as a function of pressure at the 10.5 GPa is more evident for the hexagonal phase, the adopted procedure allows for the quantification of the bulk moduli of both phases, which changes from (131 ± 1.7) to (204.7 ± 25.9) GPa and from (7.4 ± 0.3) to (14.2 ± 1.0) GPa for hexagonal and cubic phases, respectively. The putative cubic structure at the intermediate pressure region undergoes further hardening in the third region, accompanied by a simultaneous hardening of the hexagonal part of the compound.

Table 1 – Bulk modulus $B_0$ and unit cell volume $V_0$ parameters obtained from the optimization of a second-order Birch-Murnaghan equation.

| w-ZnO | I + II | III |
|---|---|---|
| $V_0$ [Å³] | 47.6 ± 0.02 | 46.4 ± 0.3 |
| $B_0$[GPa] | 131.3 ± 1.7 | 204.7 ± 25.9 |

| rs-ZnO | II | III |
|---|---|---|
| $V_0$ [Å³] | 118.75 ± 1.1 | 102.2 ± 1.6 |
| $B_0$[GPa] | 7.4 ± 0.3 | 14.2 ± 1.0 |

Figure 3 presents the photoluminescence spectra collected simultaneously with the X-ray diffraction patterns over the same pressure range (0.5 GPa ≤ P ≤ 13.5 GPa), in which five main intra-4f shell optical transitions of $Eu^{3+}$ were identified. These transitions are labeled as $^5D_0 \rightarrow {}^7F_0$, $^5D_0 \rightarrow {}^7F_1$, $^5D_0 \rightarrow {}^7F_2$, $^5D_0 \rightarrow {}^7F_3$ and $^5D_0 \rightarrow {}^7F_4$, following the Russell-Saunders[30] notation $^{2S+1}L_J$. Here, L and S represent the vectorial sum of the orbital and spin angular momenta of the individual electrons, respectively, while J corresponds to the total angular momentum. According to the main characteristics of these transitions discussed in [1], the $^5D_0 \rightarrow {}^7F_0$, should be present in a $C_{6v}$ symmetry, $^5D_0 \rightarrow {}^7F_2$, $^7F_4$ transitions are electric dipole (ED) allowed and exhibit hypersensitive and sensitive behavior, respectively. In particular, the $^5D_0 \rightarrow {}^7F_2$ hypersensitive transition means its intensity is strongly influenced by the surrounding crystal field. Consequently, it often appears as one of the most intense peaks in the emission spectrum when $Eu^{3+}$ ions are located in non-centrosymmetric environments. According to Judd–Ofelt theory, this transition reaches maximum intensity when $Eu^{3+}$ occupies low-symmetry sites lacking an inversion center[31–33]. The $^5D_0 \rightarrow {}^7F_1$ transition, on the other hand, is of magnetic dipole (MD) nature and typically less affected by the local environment. The transitions to the $^7F_1$ and $^7F_3$ levels are formally forbidden by both ED and MD selection rules[33,34], however, the observation of well-resolved emissions associated with these levels suggests that the Eu³⁺ ions occupy sites with low symmetry, where crystal field effects induce J-mixing[35].

It is observed that the emission peaks of each transition are composed of multiple components, which depend on the local symmetry[36]. With increasing pressure, the reduction in interatomic distances modifies the crystal field configuration, leading to changes in the relative intensity and energy shift of the optical emissions peaks. Moreover, the Eu³⁺ spectra are sensitive to the size of the lattice cations[37] and local chemical interactions, providing insights into the electronic structure and symmetry of the host site[38].

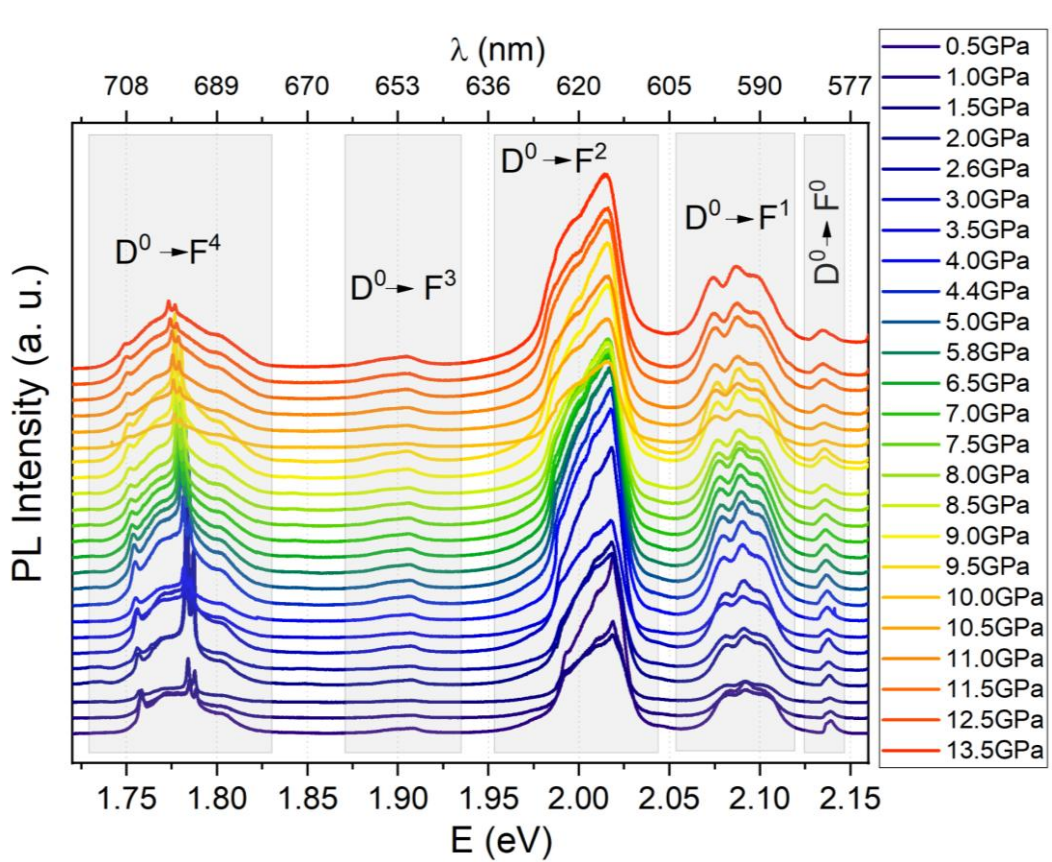

Figure 3 – Photoluminescence spectra of ZnO: Eu as a function of pressure. The highlighted Eu³⁺ ion emissions are due to the intra-4f shell $^5D_0 \to {}^7F_J$ (J = 0, 1, 2, 3, 4) optical transitions.

To the best of our knowledge, no studies have reported on the photoluminescence (PL) response of Eu³⁺ ions as a function of external pressure. Previous investigations[39,40] have focused on europium oxide ($Eu_2O_3$) under applied pressure to analyze the valence variation between the $Eu^{2+}$ and $Eu^{3+}$ states. These studies revealed a shift in the valence ratio within the pressure range of 14.0 GPa ≤ P ≤ 40.0 GPa, where $Eu^{3+}$ becomes predominant. At higher pressures, the $Eu^{2+}$ state is re-established[40]. To further investigate the optical emission behavior of $Eu^{3+}$ under varying pressure conditions, the emission of each $D_0 \rightarrow F_J$ transition was fitted by an envelope formed by the sum of contributions from individual Gaussian peak components within the spectral profile. The analysis focused on the evolution of the peak energy position and full width at half maximum (FWHM) as a function of pressure, are shown in Figures 4 (a) and (b), respectively. The inset images illustrate the normalized PL spectra, exemplified by the $^5D_0 \rightarrow {}^7F_0$ transition, highlighting the observed peak shift and broadening with increasing pressure. All emission peaks exhibited a similar trend, with a systematic shift toward lower energies at an average rate of ~ 0.40 ± 0.02 meV/GPa (Figure 4 (a)) and a corresponding increase in peak width, averaging ~ 0.55 ± 0.02 meV/GPa (Figure 4 (b)).

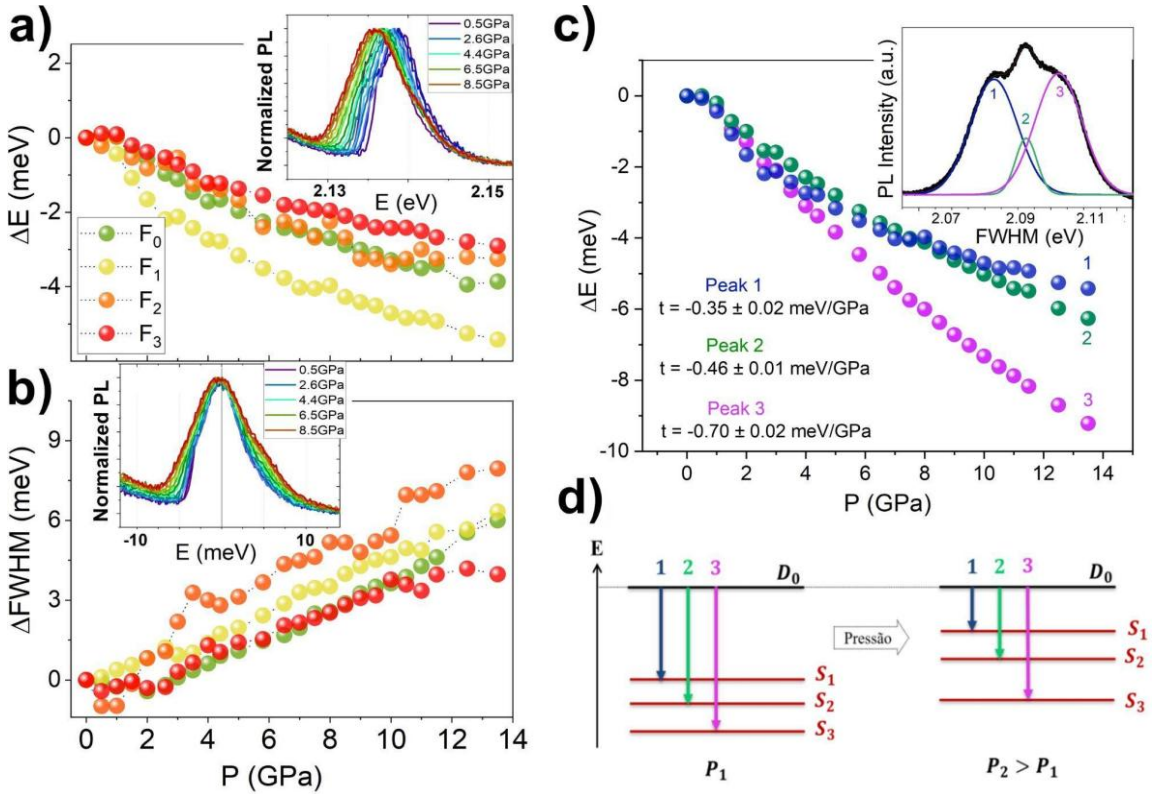

Figure 4 – Fitted values extracted from the spectra showing: (a) the shift in peak position (eV) and (b) the change in peak width (meV) as a function of pressure (GPa). The insets display a representative normalized PL spectrum ($^5D_0 \rightarrow {}^7F_0$ transition) as a function of energy, highlighting the observed variations. (c) Energy shift (meV) of the optical emissions associated with the $^5D_0 \rightarrow {}^7F_1$ transition as a function of pressure. (d)

Schematic representation of the pressure-induced changes in emissions associated with electronic transitions to the Stark sublevels ($S_1$, $S_2$, and $S_3$): all peaks exhibit a redshift (smaller arrows for $P_2 > P_1$). While the transitions $^5D_0 \rightarrow S_1$ and $S_2$ shift at similar rates ($S_1$ and $S_2$ remain closely spaced), the $^5D_0 \rightarrow S_3$ transition shows a larger shift with increasing pressure, indicating greater separation of the $S_3$ level from $S_1$ and $S_2$.

The application of hydrostatic pressure leads to a reduction in the lattice parameters of ZnO, particularly along the c-axis of the hexagonal phase, as revealed by XRD analysis. This compression promotes a diffuse structural transition toward the cubic phase. As the lattice contracts, the resulting increase in atomic interactions enhances the crystal field strength, accounting for the pressure-induced peak shifts and broadenings. Studies on the luminescence of rare-earth ions have shown that an increase in the lattice parameter leads to a redshift in PL peaks, an effect attributed to the reduction in crystal field intensity[41]. Similarly, in the sample analyzed in this study, the decrease in the lattice parameter and the consequent increase in crystal field intensity result in a shift toward lower energies. The shift in PL peaks is a significant phenomenon and can be directly related to the atomic structure of the material. In the case of ZnO, redshifts in the UV band have been observed depending on the morphology of structures synthesized via aqueous chemical growth[42] and due to a reduction in the average nanoparticle size[43]. For $Eu^{3+}$, spectral shifts toward orange and yellow emissions have also been reported[1], attributed to luminescence from higher excited states ($^5D_1, ^5D_2, ^5D_3$). This effect can be tuned by adjusting the $Eu^{3+}$ concentration in the host matrix.

To understand the broadening of the emission peaks, two main contributions should be considered: *(i)* the emergence of multiple emissions associated with the same electronic transition, due to Stark splitting of previously degenerate levels, and *(ii)* a differential shift of individual Stark components as pressure increases. The Stark effect, which causes the splitting of degenerate energy levels into multiple sublevels, arises from the interaction between an electric field and the dipole moment of the energy states[44], resulting in additional emission lines in the spectrum. To investigate this behavior, Figure 4 (c) presents the evolution of the energy corresponding to the maximum intensity of each well-resolved Stark component in the $^5D_0 \rightarrow ^7F_1$ transition as a function of pressure.

The inset shows the spectral band fitted with three Gaussian peaks. The distinct shift rates (t) of the peaks indicate that the increase in external pressure does not lead to additional emissions associated with new Stark levels, but rather modifies the spacing between the existing ones. Accordingly, the enhancement of the ZnO crystal electric field not only influences the 4f emissions of the rare-earth ion, but also affects the relative behavior of the Stark components themselves. The observation of three Stark components in the $^7F_1$ transition is attributed to the influence of a strong crystal field acting on the $Eu^{3+}$ ions[35]. A representation of the emissions corresponding to the Stark sublevels ($S_1$, $S_2$, and $S_3$) under increasing pressure for the same transition is shown in Figure 4 (d), which illustrates both the redshift and the increased energy separation between the electronic levels as a consequence of the applied hydrostatic pressure.

The PL spectra for each transition are shown in Figure 5, corresponding to transitions between the $D_0$ level and the a) $F_1$, b) $F_0$, c) $F_2$, d) $F_3$, and e) $F_4$ levels. Decomposed peaks (in gray) are also displayed for each case. These peaks correspond to transitions to Stark sublevels, as illustrated in Figure 4 (a) for the $D_0 \rightarrow F_1$ transition. The maximum number of Stark sublevel splittings for $Eu^{3+}$ follows the 2J+1 rule, given that J is an integer in all observed electronic transitions, according to the RS coupling notation. Consequently, the $D_0 \rightarrow F_4$ (J = 4) transition is expected to exhibit up to nine Stark peaks, while the $D_0 \rightarrow F_3$ (J = 3) transition can present a maximum of seven peaks, $D_0 \rightarrow F_2$ (J = 2) up to five peaks, and $D_0 \rightarrow F_1$ (J = 1) up to three peaks. Finally, for the $D_0 \rightarrow F_0$ (J = 0) transition, a single optical emission peak is expected. From Figure 5, it is observed that most transitions exhibit the maximum expected number of optical emission peaks associated with Stark sublevels, except for the $F_3$ level transition, which presents only two Stark peaks out of the seven theoretically possible. However, as mentioned before, this transition is strongly forbidden and typically occurs only when the ion is located in a chemically asymmetric environment[35]. In such cases, the mixing of electronic states induced by the local asymmetry can partially lift the selection rules, allowing the transition to occur. In the case of the $^7F_2$ transition, the splitting into three sublevels due to the Stark effect has also been previously reported[32,45]. Therefore, the presence of multiple Stark emissions provides further evidence of the strong influence of an asymmetric crystal field from the ZnO host on the rare-earth ion in this sample. Stark-split emissions have been reported as sensitive indicators of local symmetry around rare-earth ions like $Er^{3+}$, with more Stark components indicating lower symmetry in the ion's chemical environment[46].

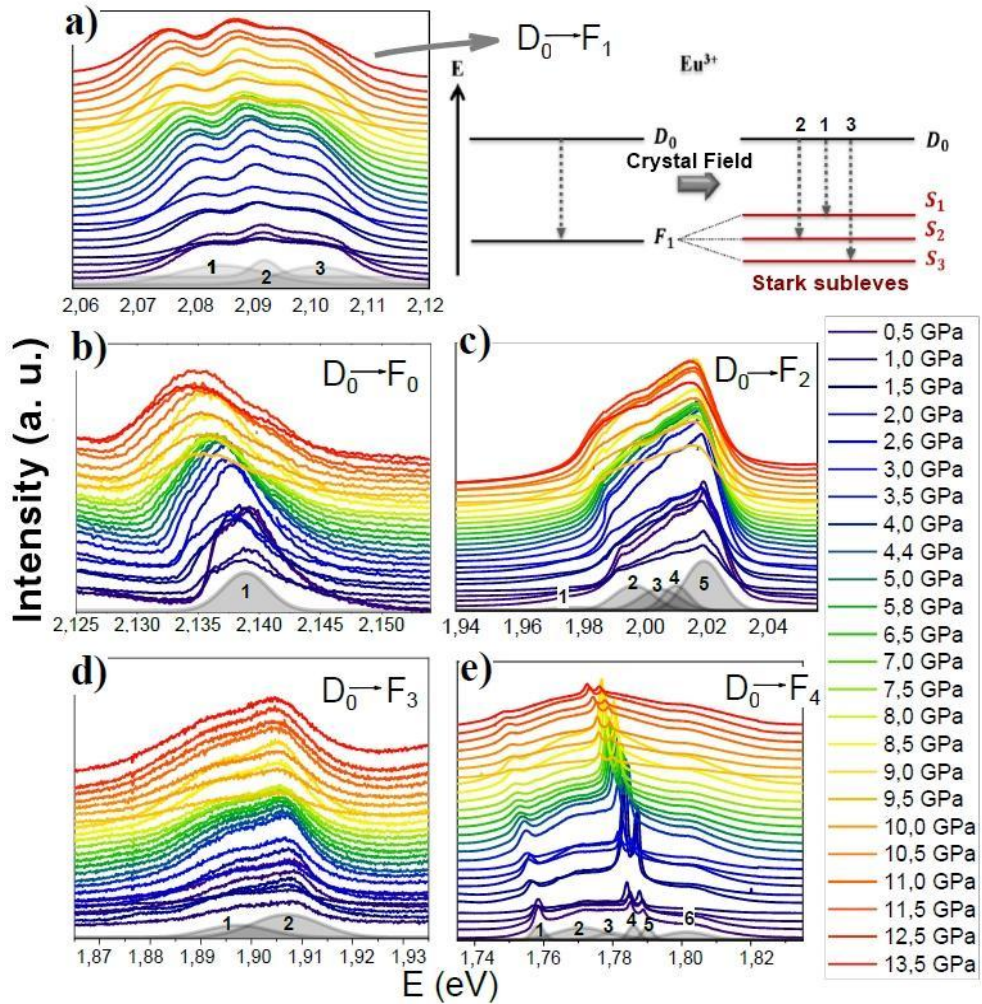

Figure 5 – PL spectra as a function of energy (eV) for the electronic transitions: a) $D_0 \rightarrow F_0$, b) $D_0 \rightarrow F_1$, c) $D_0 \rightarrow F_2$, d) $D_0 \rightarrow F_3$, and e) $D_0 \rightarrow F_4$. In a), the splitting of the $F_1$ level into three Stark sublevels ($S_1$, $S_2$, and $S_3$) due to the influence of the crystal field is depicted. The splitting in the sublevels results in three emission peaks (PL graphs in gray), each represented by arrows indicating transitions from the $D_0$ level to the corresponding sublevels, based on their energy values.

Figure 6 (a) shows the fitted PL intensities of $Eu^{3+}$ transitions $D_0 \rightarrow F_J$ as a function of applied hydrostatic pressure. The data presented in the graphs indicate that, in the pressure range between 9.5 and 10.0 GPa (highlighted region), all $Eu^{3+}$ emissions exhibit a significant decrease in PL intensity. Simultaneously, as shown in Figure 6 (b), there is a sharp increase in the relative amount of rs-ZnO for the same range, evidenced by a drop in the integrated XRD intensity ratio between the wurtzite (w-ZnO) and rocksalt (rs-ZnO) phases. The integrated intensities of the w-ZnO and rs-ZnO phases were determined by performing a Gaussian fit on each peak observed in the XRD patterns, with the areas corresponding to each phase summed accordingly. The maximum ratio is 171.4 at 0.5 GPa, while the minimum reaches ~7.0 at 12.5 and 13.5 GPa. These results indicate a pressure-induced phase transition from the hexagonal to the polymorph structure, particularly around 9.5 – 10.5 GPa. Given that the cubic structure is more symmetric than the hexagonal one (Fm-3m, n°225 and P6$_3$mc, n°186, respectively), the observed PL quenching is attributed to a growing proportion of highly symmetric regions within the material. According to the Judd-Ofelt theory previously mentioned[31,47], this behavior is expected: in highly symmetric environments, 4f–4f transitions are parity-forbidden and predominantly occur via magnetic dipole mechanisms, which are much weaker. As the local symmetry decreases, the crystal field becomes more distorted, increasing the mixing of opposite parity states and allowing electric dipole transitions to become partially allowed, thereby enhancing the luminescence efficiency of rare-earth ions such as $Eu^{3+}$. As shown in Figure 6 (a), PL intensity exhibits a renewed increase when pressure exceeds 10 GPa, suggesting a possible structural or electronic reconfiguration. This behavior is likely associated with the growing contribution of amorphous regions derived from the wurtzite phase, which becomes more pronounced at higher pressures. Figure 6 (c) presents the FWHM evolution of w-ZnO diffraction peaks (w-(100), w-(002), w-(101)) above 5.8 GPa. The inset shows normalized XRD intensity, highlighting FWHM broadening. From this pressure on, the FWHM increases at a rate of approximately ~0.005°/GPa, although the trend is not strictly linear. Broader diffraction peaks suggest increased structural disorder and amorphization of the wurtzite phase. At 13.5 GPa, the highest pressure investigated, the wurtzite structure is not fully converted to the cubic phase, in contrast with some reports for undoped ZnO[15,28]. However, the continuous broadening of the w-ZnO peaks confirms progressive amorphization under pressure. Notably, the discontinuities in the PL and XRD data occur within the same pressure range, reinforcing the correlation between structural transformations in the host matrix and variations in $Eu^{3+}$ emission. These findings demonstrate that structural symmetry and crystallinity play a critical role in governing the optical behavior of $Eu^{3+}$-doped ZnO. The observed Stark splitting and emission quenching reflect the sensitivity of the rare-earth luminescence to local crystal field variations, which are strongly influenced by pressure-induced structural changes.

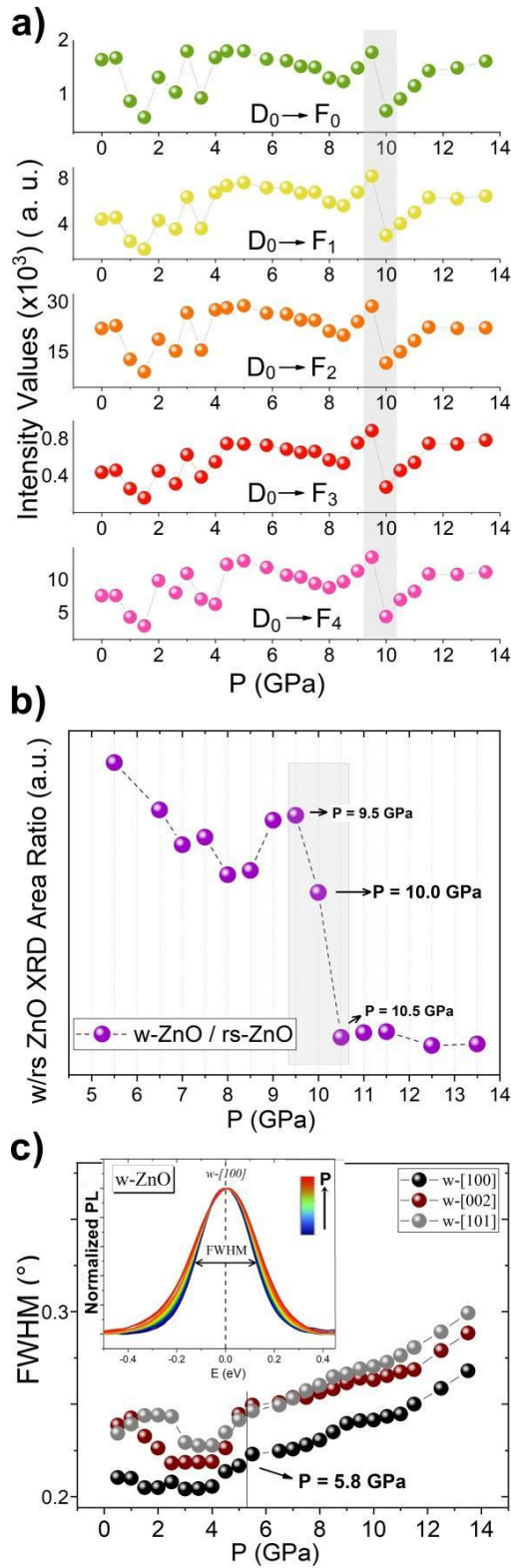

Figure 6 – a) Peak intensities as a function of pressure for the $D_0 \rightarrow F_J$ transitions. The highlighted region indicates a sharp drop in emission intensity for all transitions at $P$ = 10 GPa. b) Ratio between the integrated XRD area of w-ZnO and rs-ZnO phases, showing an abrupt decrease between $P$ = 9.5 GPa and $P$ = 10.5 GPa. c) FWHM of the w-ZnO peaks as a function of pressure (P) shows a noticeable increase above 5.8 GPa. The inset presents the normalized XRD intensity, highlighting the broadening of the FWHM for the w-(100) peak.

To provide a detailed understanding of local structural and electronic effects, we performed first-principles simulations to explore the behavior of $Eu^{3+}$ ions in distinct lattice environments. To model interstitially Eu-doped ZnO structures, we constructed a 4×4×2 supercell for the hexagonal phase and a 2×2×2 supercell for the cubic phase. In each case, a single Eu atom was introduced, corresponding to a doping concentration of approximately 3% relative to the total number of Zn atoms. The effect of hydrostatic pressure was incorporated by fixing the lattice parameters to the values obtained from Rietveld refinement at each pressure point. Subsequently, only the internal atomic coordinates were relaxed, allowing the system to adjust to the imposed volume constraints while capturing pressure-induced changes in local atomic environments.

Figure 7 presents the projected density of states (PDOS) for Eu-doped ZnO in both the hexagonal [(a1)–(c1)] and cubic [(d1)–(f1)] phases at selected pressures (0, 10, and 13.5 GPa), highlighting the contributions from Zn-3d, O-2p, and Eu-4f orbitals. In both phases, the Zn-3d and O-2p orbitals form a broad valence band extending significantly below the Fermi energy, while the Eu-4f states appear as sharp localized peaks, characteristic of their weak hybridization with the host lattice. The Zn-3d states are primarily located in the deeper portion of the valence band, whereas the O-2p states dominate the upper valence region, closer to the Fermi level. However, strong hybridization between O-2p and Zn-3d orbitals is evident across the entire valence band, shaping the overall electronic structure of the host matrix and modulating the local environment of the Eu-4f states. Notably, the extent of Zn-O hybridization appears slightly more delocalized in the cubic phase compared to the hexagonal one, likely reflecting the higher coordination and symmetry of the rocksalt lattice.

For the hexagonal phase, the Eu-4f states at ambient pressure Figure 7a1) are narrowly concentrated just below the Fermi level, with slight orbital-dependent shifts, as shown in Figure 7a2. As pressure increases to 10 GPa (Figure 7b2), these peaks shift toward lower energies and broaden, accompanied by increased separation between the 4f orbitals. By 13.5 GPa, the orbital splitting becomes more pronounced (Figure 7c2), revealing an enhancement of the crystal field interaction as the lattice compresses. This progression reflects the sensitivity of the $Eu^{3+}$ local environment to pressure and agrees with the experimental observation of Stark splitting in the photoluminescence spectra.

In contrast, the cubic phase exhibits a different behavior. At 0 GPa (Figure 7d1), the Eu-4f states remain sharply defined, but the orbital decomposition reveals a nearly degenerate group of six components (Figure 7d2). Only the $4f_{xyz}$ orbital appears distinctly separated from the others, which cluster closely in energy. This near-degeneracy persists at higher pressures (Figures 7e2 and 7f2), even though the entire 4f manifold shifts to lower energies with increasing pressure. The persistence of orbital degeneracy in the cubic phase is a consequence of its higher lattice symmetry (Fm–3m), which imposes stricter constraints on the crystal field splitting of the 4f levels. In such a symmetric environment, the crystal field potential lacks sufficient anisotropy to strongly lift the degeneracy of the Eu-4f states, except for the $4f_{xyz}$ orbital, which appears to be more sensitive to subtle symmetry-breaking interactions or strain-induced distortions.

This contrast between phases is crucial to understanding the optical behavior of $Eu^{3+}$ in ZnO under pressure. The hexagonal phase, with its lower symmetry, permits a more pronounced and continuous splitting of the 4f orbitals, which becomes progressively stronger with compression. This leads to observable Stark splitting and broadening in the emission lines. On the other hand, the cubic phase maintains a more degenerate 4f structure even under pressure, consistent with the experimentally observed quenching of photoluminescence as the system transitions to a more symmetric phase. Therefore, the evolution of the Eu-4f states in the PDOS not only confirms the phase transition but also provides a microscopic explanation for the pressure-tunable optical properties of Eu-doped ZnO.

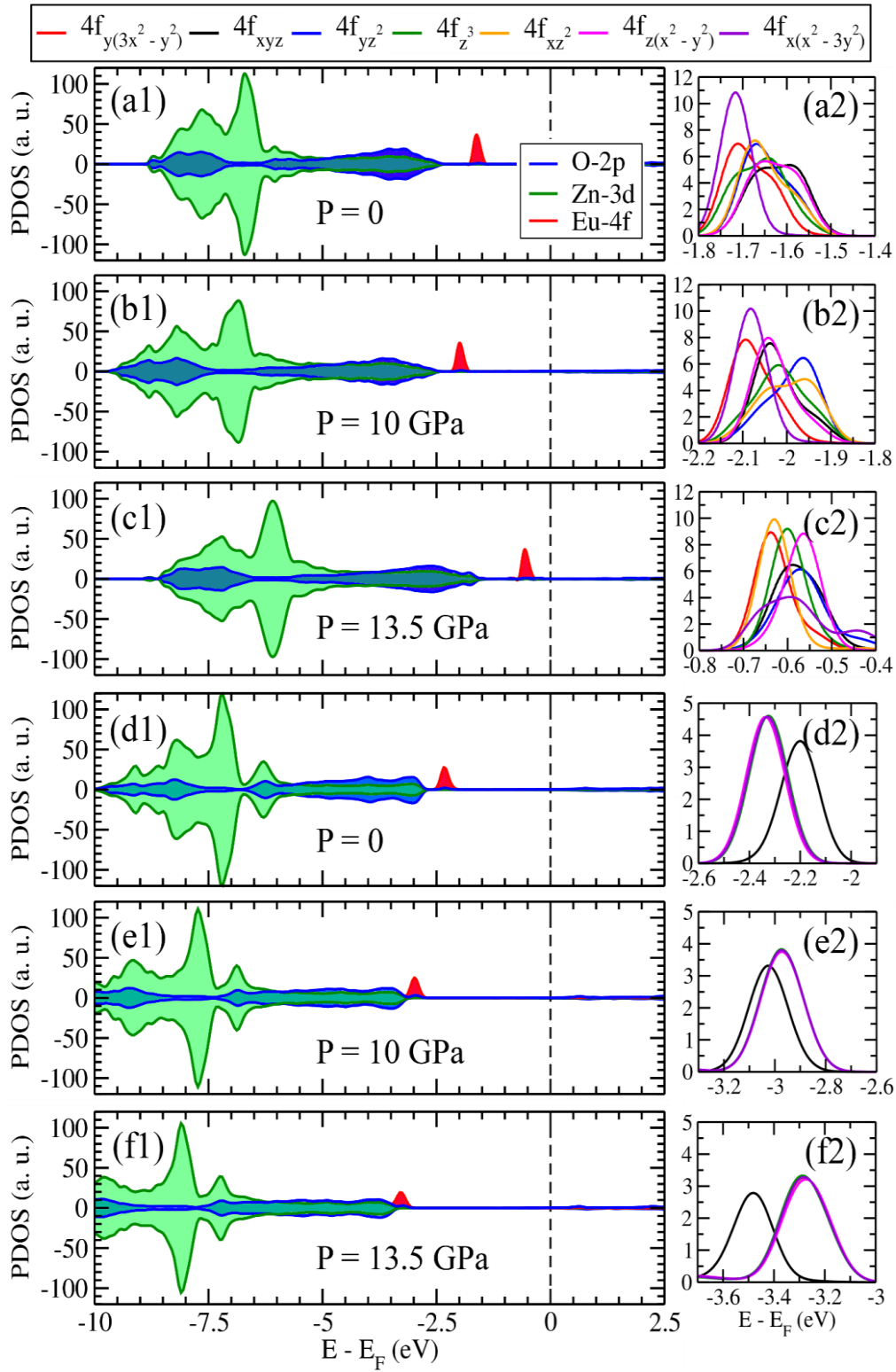

Figure 7 – Projected density of states (PDOS) for Eu-doped ZnO in the hexagonal [(a1)–(c1)] and cubic [(d1)–(f1)] phases at selected pressures: 0, 10, and 13.5 GPa. Each panel shows the total PDOS resolved into Zn-3d (green), O-2p (blue), and Eu-4f (red) contributions. The corresponding orbital decomposition of the Eu-4f states is shown in panels (a2)–(f2), revealing the pressure-induced evolution of the 4f-level splitting.

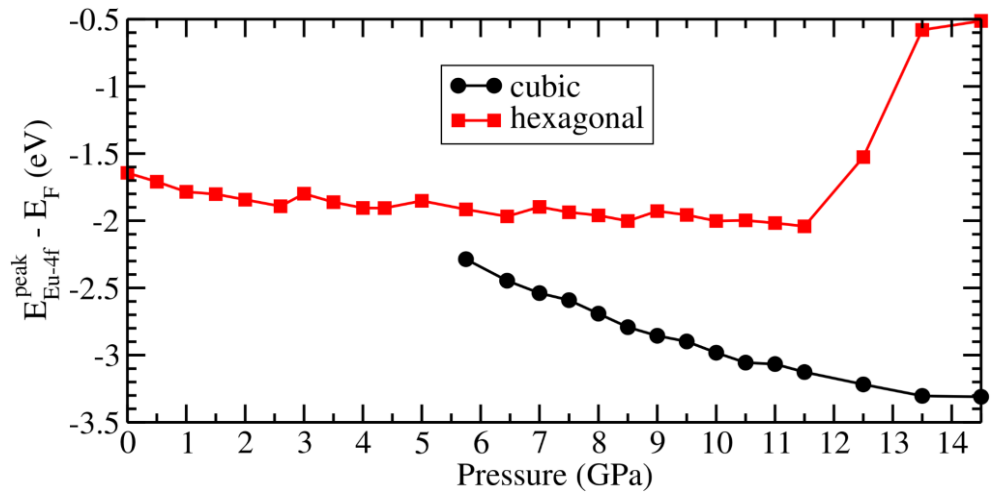

Figure 8 – Pressure dependence of the Eu-4f peak energy relative to the Fermi level ($E_F$) for Eu-doped ZnO in hexagonal (red squares) and cubic (black circles) phases, obtained from DFT calculations.

Figure 8 presents the evolution of the energy position of the Eu-4f peak relative to the Fermi level as a function of pressure for both hexagonal and cubic ZnO:Eu structures. In the hexagonal phase, the Eu-4f peak energy remains nearly constant across the 0–11.5 GPa pressure range, exhibiting a minor initial decrease upon pressure application and subsequently fluctuating only slightly around –1.9 eV. This stability reflects the relatively modest influence of pressure on the local electronic environment of the $Eu^{3+}$ ion within the wurtzite lattice. However, a sharp discontinuity occurs beyond 11.5 GPa, where the Eu-4f peak abruptly shifts toward the Fermi level, reaching values near –0.5 eV at 13.5 and 14.5 GPa. This sudden shift coincides with the experimentally observed structural transformation and indicates a critical change in the crystal field environment. In contrast, the cubic phase exhibits a markedly different trend. As pressure increases from 5.75 to 14.5 GPa, the Eu-4f peak systematically shifts to lower energies concerning the Fermi level, following a nearly linear trend. This monotonic downward shift reflects a progressive stabilization of the Eu-4f states under compression, likely due to the increased orbital overlap between Eu ions and neighboring oxygen atoms within the denser rocksalt lattice. The continuous nature of this trend contrasts with the abrupt behavior seen in the hexagonal case and suggests that, once formed, the cubic structure provides a more uniform and symmetric crystal field environment for the Eu ion.

These contrasting behaviors highlight the sensitivity of Eu-4f states to structural symmetry and local coordination. The abrupt jump in the hexagonal case reflects a destabilization of the 4f levels due to increasing disorder or a transition to a metastable environment, while the gradual shift in the cubic phase points to a consistent evolution of the crystal field with pressure. Importantly, these electronic structure changes correlate with the experimentally observed pressure-dependent photoluminescence behavior, which the stabilization and shift of the Eu-4f levels in the cubic phase align with the redshift and quenching of emission, while the discontinuous change in the hexagonal phase corresponds to the onset of structural transformation and the partial recovery of luminescence at higher pressures. This comparison reinforces the role of local symmetry and structural phase in governing the optical properties of Eu-doped ZnO under pressure conditions.

## Conclusions

In summary, we have demonstrated the strong sensitivity of $Eu^{3+}$ photoluminescence to pressure-induced structural transformations in ZnO lattice. Synchrotron XRD measurements revealed a gradual phase transition from the hexagonal wurtzite to the cubic rocksalt structure starting at ~5.8 GPa, with the cubic phase becoming predominant above 10 GPa. This transformation leads to a pronounced quenching of $Eu^{3+}$ optical emissions, consistent with the enhanced crystal symmetry of the rocksalt phase. Partial photoluminescence recovery at higher pressures suggests the emergence of disordered or amorphous regions. First-principles DFT calculations helped to clarify the underlying electronic origins of the experimentally observed optical behavior of Eu-doped ZnO under pressure. The computed electronic structure confirmed that compression stabilizes and modulates the Eu-4f levels, with distinct orbital shifts observed in hexagonal and cubic phases. In particular, while the hexagonal structure becomes unstable under higher pressures, the cubic phase shows a steady stabilization and redshift of the Eu-4f states, which directly matches the observed optical behavior. These findings establish hydrostatic pressure as an effective tool for engineering the structural and optical properties of rare-earth-doped semiconductors through symmetry-driven modulation of their electronic environments.

## Acknowledgements

This research used facilities of the Brazilian Synchrotron Light Laboratory (LNLS), part of the Brazilian Center for Research in Energy and Materials (CNPEM), a private non-profit organization under the supervision of the Brazilian Ministry for Science, Technology, and Innovations (MCTI). The EMA beamline staff is acknowledged for the assistance during the experiments under proposal number 20221248. ADR, CIP, and MPFG acknowledge FAPESP (#2021/13974-0, #2024/02854-1) and CNPq (#310819/2023-7, #312254/2023-7). I.S.S.d.O. acknowledges financial support from FAPEMIG, and computer time from LCC-UFLA and CENAPAD-SP.